\input{aipcheck}

\documentclass[
    ,final            
  ]
  {aipproc}

\layoutstyle{6x9}

\begin{document}

\title{Bowen blend echo-tomography of low mass X-ray binaries}


\author{T.~Mu\~noz-Darias}{
  address={Instituto de Astrof\'\i{}sica de Canarias, 38200 La Laguna, 
  Tenerife, Spain}
}
\author{I.~G.~Mart\'\i{}nez-Pais}{
  address={Instituto de Astrof\'\i{}sica de Canarias, 38200 La Laguna, 
  Tenerife, Spain}
}
\author{J.~Casares}{
  address={Instituto de Astrof\'\i{}sica de Canarias, 38200 La Laguna, 
  Tenerife, Spain}
}
\author{R.~Cornelisse}{
  address={Instituto de Astrof\'\i{}sica de Canarias, 38200 La Laguna, 
  Tenerife, Spain}
}
\author{V.~S.~Dhillon}{
  address={Dept. of Physics \& Astronomy, Univ. of Sheffield, Sheffield  
  S3 7RH, UK}
}
\author{T.~R.~Marsh}{
  address={Dept. of Physics, Univ. of Warwick, Coventry CV4 7AL, UK}
}

\author{D.~Steeghs}{
  address={Harvard-Smithsonian Center for Astrophysics, Cambridge  
  MA 02138, USA}
}
\author{K.~O'Brien}
{address={European Southern Observatory, Casilla 19001, Santiago 19, Chile}}
\author{P.~A.~Charles}{
 address={South Africa astronomical observatory, PO Box 9, observatory, South Africa}
} 
 \author{M.~Still}{
  address={South Africa astronomical observatory, PO Box 9, observatory, South Africa}
}

\begin{abstract}
 We present simultaneous high time resolution (1-10 Hz) X-ray and optical observations of the persistent LMXBs Sco X-1 and V801 Ara(=4U 1636-536). In the case of Sco X-1 we find that the Bowen/HeII emission lags the X-ray light-curves with a light travel time of $\sim 11-16$s which is consistent with reprocessing in the donor star. We also present the detection of three correlated X-ray/optical bursts in V801 ara. Although this latter project is still in progress our preliminary results obtained by subtracting the Continuum light-curve from the Bowen/HeII data provide evidence of orbital phase dependent echoes from the companion star.
\end{abstract}

\maketitle


\section{Introduction} \label{introduction}
\noindent Low mass X-ray binaries (LMXBs) are interacting binaries containing a low mass donor star
transferring matter onto a neutron star (NS) or black hole.
Mass accretion takes place through an accretion disc, and with temperatures approaching $\sim 10^8$ K,
such systems are strong X-ray sources. The mass transfer rate supplied by the donor star, $\dot{M}_2$, is driven 
by the binary/donor evolution and for $\dot{M}_2>\dot{M}_{crit}\sim 10^{-9} M_{\odot} yr^{-1}$ results in persistently 
bright X-ray sources. In these binaries optical emission is dominated by reprocessing of the powerful X-ray luminosity in the gas around the compact object which usually swamp  the spectroscopic features of the weak companion stars. 
In this scenario, dynamical studies have classically been
restricted to the analysis of X-ray transients during quiescence, where the intrinsic
luminosity of the donor dominates the optical spectrum of the binary.\\
Fortunately, this situation has recently changed thanks to the discovery
of the narrow emission components arising from the donor star in the prototypical persistent
LMXB Sco X-1 \cite{stee02}. High resolution spectroscopy revealed many narrow high-excitation
emission lines, the most prominent associated with HeII $\lambda4686$ and NIII $\lambda\lambda$4634-41 /
CIII $\lambda\lambda$4647-50 at the core of the broad Bowen blend. In particular, the NIII lines are powered by fluorescence resonance through
cascade recombination which initially requires EUV seed photons of HeII
Ly${\alpha}$.  These very narrow (FWHM $\leq 50$km s$^{-1}$) components move in antiphase with
respect to the wings of HeII $\lambda$4686, which
approximately trace the motion of the compact star. Both properties
(narrowness and phase offset) imply that these components originate in
the irradiated face of the donor star. We repeated this experiment with success in other 8 sources (see e.g. \cite{C04}) 
which imply that mass donor star irradiation and subsequent Bowen blend fluorescence is a common occurrence in persistent LMXBs.
\subsection{Echo-Tomography}
Echo-tomography is an indirect imaging technique which uses time delays
between X-ray and UV/optical light-curves as a function of orbital phase
in order to map the reprocessing sites in a binary \cite{obrien02}.
The optical light curve can be simulated by the
convolution of the (source) X-ray light curve with a transfer function which
encodes information about the geometry and visibility of the reprocessing
regions. The transfer function~(TF) quantifies the binary response to the
irradiated flux as a function of the lag time and it is expected to have two main components:
the accretion disc and the donor star. The latter is strongly dependent on
the inclination angle, binary separation and mass ratio and, therefore,
can be used to set tight constraints on these fundamental parameters.
Successful echo-tomography experiments have been performed on several X-ray
active LMXBs using X-ray and broad-band UV/optical light-curves. The results
indicate that the reprocessed flux is dominated by the large
contribution of the accretion disc (e.g. \cite{hynes98},\cite{obrien02}) which dilutes the reprocessed signal arising from the companion. Only \cite{midd81} and \cite{dav75} have reported the detection of reprocessed pulsations in the X-ray binaries 4U 1626-67 and Her X-1 respectively. More recently, \cite{hynes06} have reported possible evidence for a reprocessed signal from the companion 
through the detection of orbital phase dependent echoes in EXO 0748-676.\\

\section{Bowen blend echo-tomography of LMXBs}
Based on our previous works where we detect reprocessed Bowen emission from the donor star in many LMXBs, we developed the idea of Bowen blend echo-tomography. To do this, we manufactured two narrow
(FWHM =100 \AA) interference filters 
centered at the Bowen and HeII lines ($\lambda_{\rm eff}$=4660\AA) and a featureless region at $\lambda_{\rm eff}$=6000\AA.
These filters were built to be used with ULTRACAM, a triple-beam CCD camera which uses two dichroics to split the
light into 3 spectral ranges: Blue ($<\lambda$3900),
Green($\lambda\lambda$3900-5400) and Red ($>\lambda$5400).
It uses frame transfer 1024x1024 E2V CCDs which are
continuously read out, and are capable of time resolution down to 2 milliseconds by
reading only small selected windows (see \cite{Dhi07} for details). The combination of ULTRACAM and the narrow band filters 
allows us to obtain continuum-subtracted light-curves of the
high excitation lines where the response of the donor's contribution is amplified by suppressing most of the background
continuum light, which is associated with the disc. 
As a first step, we decided to undertake an echo-tomography campaign on the brightest LMXB of all,
Sco X-1 which is also the prototypical $Z-source$. Here we also present preliminary results of a new echo-tomography campaign on the X-ray burster V801 Ara. 
\subsection{Echoes from the companion in Sco X-1}
Simultaneous X-ray and optical data of Sco X-1 were obtained on the nights
of 17-19 May 2004.  
For the X-ray observations we used the Proportional Counter Array (PCA) onboard the Rossi X-ray Timing Explorer (RXTE) satellite. 
The STANDARD-1 mode, with a time resolution of 0.125s, was used for the variability analysis and the STANDARD-2 mode data, with a time resolution of 16s, was used for the spectral analysis.
Sco X-1 was observed during 15 RXTE windows of 16-31 minutes length spread over the three nights and yielding 20 ks of data. 
About 80\% of these observations were also covered with simultaneous optical data.\\
  \begin{figure}
\includegraphics[height=10cm,width=10cm]{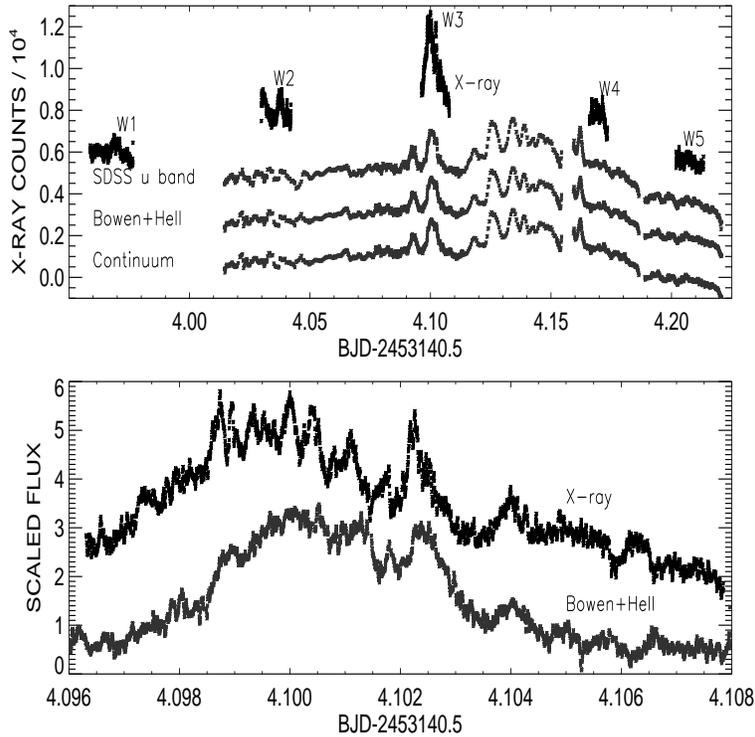}
\caption{Upper panel: Simultaneous RXTE and ULTRACAM (SDSS $u'$, Bowen+HeII and Continuum) observations of Sco X-1/V818 Sco obtained on the night of 18 May 2004. Lower panel: Zoom of W3 showing a 1000s stream of X-ray (above) and Bowen+HeII non continuum subtracted (below) data. The variability appears clearly correlated.}
\label{data}
\end{figure}
The optical data were obtained with ULTRACAM at the Cassegrain focus of the 4.2m William Herschel Telescope (WHT) at La Palma. We used our interference filters in the green (Bowen+HeII) and red (Continuum) arms of the intruments. A standard SDSS $u'(\lambda3543$ \AA) filter was also mounted in the blue channel.
The images were reduced using the ULTRACAM pipeline software with bias subtraction and flatfielding. 
The exposure time was initially set to 0.1s but was increased to 0.25s depending on weather conditions.\\

\begin{figure}
\includegraphics[height=12cm,width=12cm]{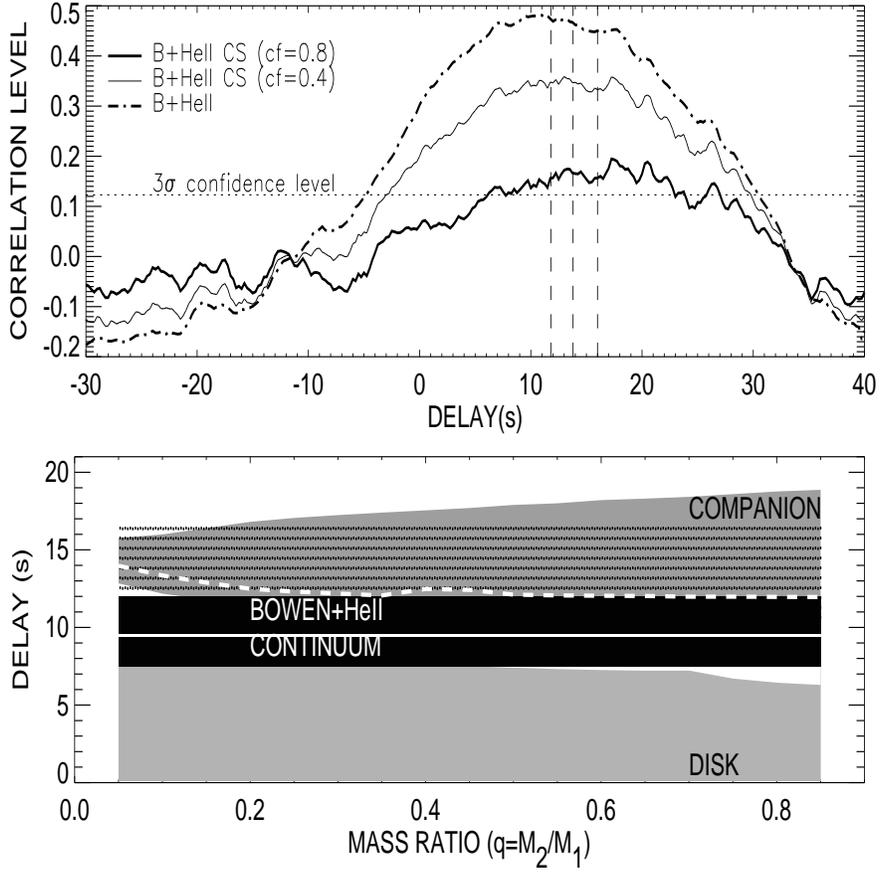}
\caption{Upper panel: ICF for the selected 3min block of W3. The ICFs have been computed using the X-ray data and a Bowen+HeII light curve obtained by subtracting from the green channel the Continuum data scaled by a factor $cf=0$ (i. e. no subtraction), 0.4 and 0.8 . The dotted line shows the $3\sigma$ confidence level and the dashed vertical lines time-lags of $11.8\pm0.1$s, $13.3\pm0.1$s and  $16.0\pm0.3$s respectively. Lower panel: Expected range of delays for both the companion and the accretion disc considering $i=50^{\circ}$, $M_1=1.4M_{\odot}$ and the orbital parameters of Sco X-1. The black region shows the $1\sigma$ delays obtained for the Bowen+HeII, and Continuum data. The shadowed region represents the range of delays obtained by subtracting the Continuum from the Bowen+HeII light-curve. The white dashed line is the most probable delay for the companion according to \cite{MD05} and using $\alpha=6^{\circ}$.}
\label{result}
\end{figure}

In order to look for correlated variability we performed a comprehensive cross-correlation analysis by using the entire data set in blocks from 2 min to 30 min (i.e. one RXTE window).To do this, we used a modified version of the ICF (Interpolation Correlation Function) method which is explained in great detail by \cite{gask87}. Significant ICF peaks were found only on the night of 18 May, when the system stayed at the Flaring Branch and the X-ray flux rose up to $1.3 \times 10^4$ counts s$^{-1}$. During this night the system shows long episodes of flaring activity (fig. \ref{data} upper panel) and correlated variability between the X-ray data and the optical lightcurve was observed (fig. \ref{data} lower panel). We obtain highly significant correlation peaks in the three considered filters by choosing different data blocks during W3 (see fig \ref{data}), especially during the second part of this window where strong correlated variability is present. W3 is centered at orbital phase 0.52, i.e. very close to the superior conjunction of the donor star, when the irradiated face presents its largest visibility with respect to us. In many cases we find ICF time-lags of about 11s between the X-ray and Bowen+HeII light-curves but we finally selected for a detailed analysis a 3 min block of the W3 where the correlation peaks are particularly significant. Table \ref{g_res} lists the ICF results obtained for the 3 bands considered (i.e. SDSS $u'$, Bowen+HeII and Continuum).\\
With the aim of amplifying the signal from the donor we subtracted the Red (Continuum) from the Green(Bowen+HeII) channels. However, the Continuum is also strongly correlated with the X-ray emission, and hence, the subtraction process reduces the correlation level of the signal and add noise. In order to determine the optimal amount of continuum to be subtracted, we computed several ICFs after subtracting a fraction $cf$ (with $cf$ in the range 0-1) of the red continuum from the Bowen+HeII light curve. We find that the ICF peak of the continuum subtracted lightcurve (Bowen+HeII CS) smoothly shifts to longer delays for larger $cf$ values, as expected if more disc contribution is subtracted (see upper panel in fig \ref{result}). We obtain a delay in the range $14.3-16.3$s for $cf=0.6-0.8$, when $\sim60\%$ of the continuum light is subtracted. For higher $cf$ values the ICF are nosier and the peaks are not significant.\\
\begin{table}
\caption{TF and ICF  analysis for W3}
\label{g_res}
\begin{tabular}{c | c c c}
\hline
 &  ICF$\pm1\sigma$(s) & $\tau_0\pm1\sigma$(s) & $\Delta\tau_0\pm1\sigma$(s)\\
\hline
Bowen+HeII CS & 14.3-16.3  & $13.5\pm3.0$&$8.5\pm3.5$ \\
Bowen+HeII &$11.8\pm0.1$  & $10.75\pm1.25$& $6.25\pm1.5$\\
SDSS $u'$ &$9.5\pm0.7$ &$9.0\pm1.5$&$5.5\pm2.0$ \\
Continuum&$9.0\pm0.7$ &$8.5\pm1.0$ &$5.5\pm1.5$ \\
\hline
\end{tabular}
\end{table}

As a second step in the data analysis we fitted the ULTRACAM light-curves with the results of convolving the RXTE data with a set of synthetic transfer functions. For simplicity we have used Gaussian TFs with two free parameters: the mean delay ($\tau_0$) and the standard deviation of the delay ($\Delta\tau$):
\begin{equation}
\label{tf}
TF(\tau)=A e^{\frac{1}{2}(\frac{\tau-\tau_0}{\Delta\tau})^2}
\end{equation}
where A is a normalization constant. We have applied this technique to all data sets and find $\tau_0$ values completely consistent with those obtained in the cross-correlation analysis. A summary of the results is presented in Table \ref{g_res}.\\

Figure \ref{result} (lower panel) shows the range of expected time-delays for both the companion and the accretion disc as a function of the mass ratio ($q=M_2/M_1$). 
The time-lags of $\sim 10-11$s obtained for the Bowen+HeII data during W3 lies at the lower edge of the companion region, and shows that part of this emission must arise from reprocessing on the donor star. Regarding the continuum subtracted data, the TF method yields a time-lag of $13.5\pm3.0$s which is perfectly accommodated in the companion region (fig. \ref{result}, lower panel), while the ICF method gives a narrower time-lag range ($14.3-16.3$s). This is larger than expected for the maximum response of the companion and suggests that either the mass of the NS or the orbital inclination are higher than the assumed values.
In addition the Continuum data shows delays consistent with reprocessing in the accretion disc whereas the $u'$ band present larger delays. This is probably due to the presence of many high excitation lines arising from the companion in the spectral range covered by this filter. We note that correlated variability is also found during W5 (fig. \ref{data}) when the system shows less activity. In this case we find delays consistent with reprocessing in an accretion disc with a radial temperature profile. (see \cite{MD07} for details)

\subsection{Phase-dependent echoes from the companion star in 4U~1636-536/V801 Ara}
\begin{figure}
\includegraphics[height=8cm,width=16cm]{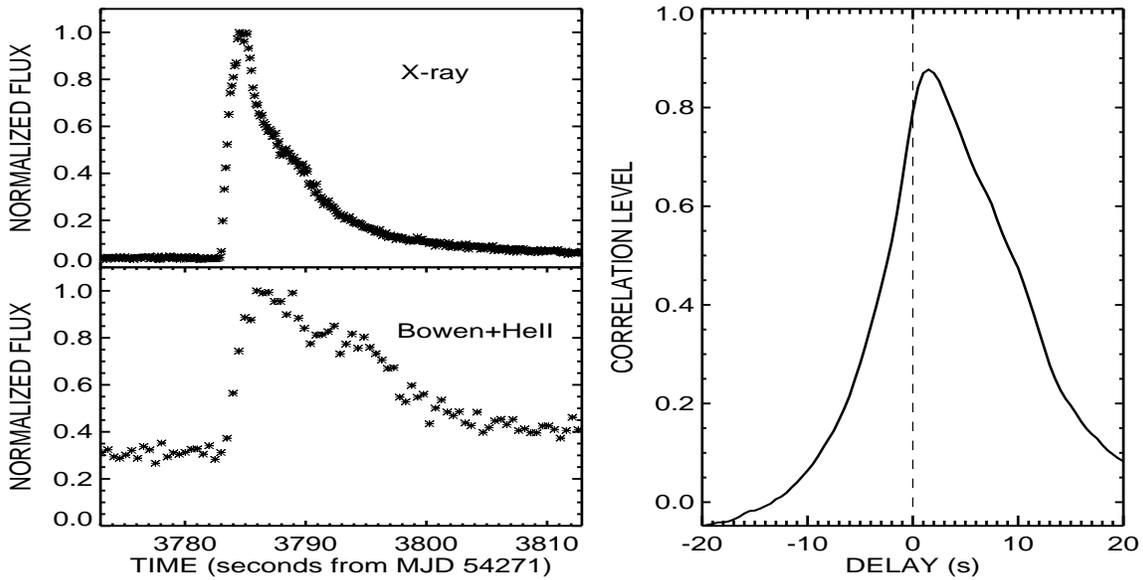}
\caption{Left panel:40s streams of X-ray (above) and Bowen+HeII (below) data around Burst-1. The smearing in the shape of the optical event is clearly seen in the figure. Righ panel: ICF obtained through the left panel lightcurves. A clear peak centered at $1.73\pm0.05$ is found.}
\label{burst1}
\end{figure}

\begin{figure}
\includegraphics[height=8cm,width=14cm]{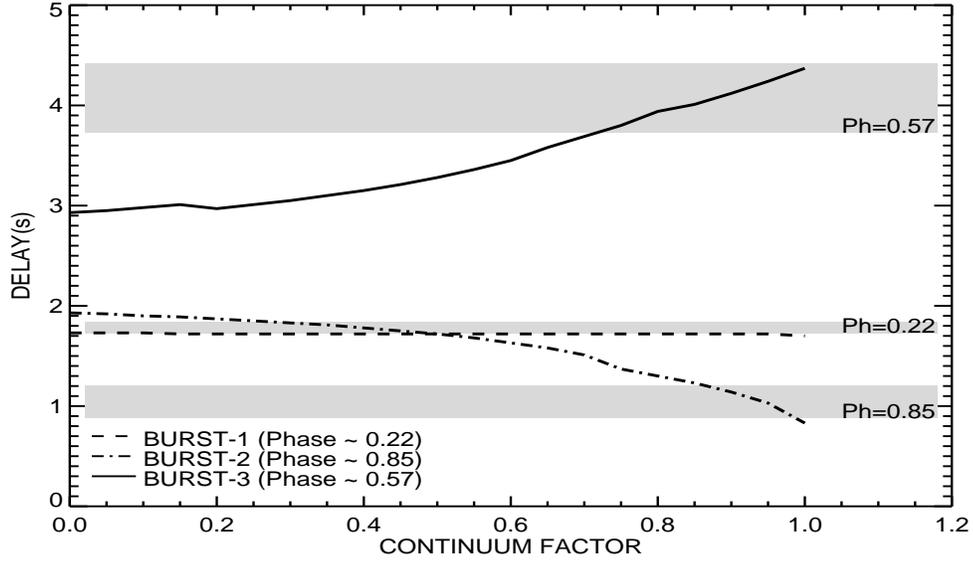}
\caption{Time-lag vs Continuum subtraction factor for the three burst detected. The grey regions show the expected range of delays for the companion assuming $q=0.265$ (\cite{C06}), $M_{NS}=1.4M_{\odot}$, disc flaring angle $(\alpha) = 6^{\circ}$, orbital inclination in the range 40-60$^{\circ}$ and the orbital phase corresponding to each burst.}
\label{rburst}
\end{figure}

Encouraged by the successful echo-tomography work on Sco X-1 we  decided to undertake a new campaign on V801 Ara where reprocessed Bowen emission from the companion has been reported \cite{C06}. This system is also a well-known X-ray burster which offers the opportunity of exploiting burst events to detect reprocessed signals from the companion at different orbital phases.
We observed V801 Ara on 19-20 June 2007 with ULTRACAM attached to the Very Large Telescope (VLT) at Cerro Paranal (Chile). We used the same instrument set up as for the Sco X-1 campaign. We used a time resolution of 0.5s, although it was increased up to 1s during the 2nd night when weather conditions decreased.  We also obtained simultaneous RXTE data covering the ground based campaing. Here we present a preliminary ULTRACAM-pipeline data reduction together with the STANDARD-1 RXTE data set at $T_{exp}=0.125$s.\\ 
Three simultaneous X-ray/optical Burst were detected in the course of the 2 nights (see fig.\ref{burst1} left panel). Table \ref{cburst} lists the times of the bursts and the orbital phase at which the events occured according to the ephemerides in \cite{C06}.\\
\begin{table}
\caption{Correlated X-ray/optical Type I burst in V801 Ara}
\label{cburst}
\begin{tabular}{c | c c c c c}
\hline
 BURST &  TIME (MJD) & $\varphi_{orbital}$ & ICF-$u'$ band(s) & ICF-Bowen+HeII(s) &ICF-Continuum(s)\\
\hline
Burst-1 & 54271.04381 & 0.22 & $1.45\pm0.02$  & $1.73\pm0.05$   & $1.73\pm0.05$ \\
Burst-2 & 54272.09179 & 0.85 & $1.75\pm0.08$  & $1.93\pm0.14$  & $2.17\pm0.08$\\
Burst-3 & 54272.20536 & 0.57 & $2.51\pm0.06$  & $2.93\pm0.06$   & $2.33\pm0.05$ \\
\hline
\end{tabular}
\end{table}
ICF functions were computed in the same way as in Sco X-1 by selecting 200s blocks of data around each burst. We find in all the cases very clear ICF peaks centered at positives delays (i.e optical event occurs after X-ray event). Figure \ref{burst1} shows the X-ray and Bowen+HeII light-curves during Burst-1 together with the resulting ICF function (right panel) whereas table \ref{cburst} lists the obtained time lags for all bands considered. We measure longer delays during Burst-3, at $\varphi_{orb}\sim0.57$, when the visibilily of the companion is larger. Furthermore, similar as for Sco X-1, the delay obtained for the Bowen+HeII filter is longer than those observed in the other bands.\\
In order  to amplify the signal from the donor we subtracted the Continuum emission scaled by a continuum factor ($cf$) from the Bowen+HeII lightcurve. Figure \ref{rburst} shows the delays obtained for each burst vs $cf$. We also mark (grey regions) the expected range of delays for the companion (following \cite{MD05}) by considering an orientative set of orbital paramenters and the orbital phase at which the events were detected. In the three cases the time-lags measured are consistent with reprocessing on the companion star. 
\section{CONCLUSIONS}
Following the discovery of narrow high excitation emission lines arising from the companion star in many LMXBs we are undertaking several Bowen blend echo-tomography experiments using ULTRACAM+narrow band filters and RXTE. Our first work on the prototypical LMXB Sco X-1 presents the detection of echoes consistent with reprocessing on the companion star. Correlated variability was found when the system stayed at the Flaring Branch and displayed high and low frequency variability \cite{MD07}. Preliminary results of a new campaign on the X-ray burster V801 Ara yield evidence of orbital-phase dependent echoes from the companion during the three burst which were detected. This latter work is still in progress and we are currently refining our results by fitting time-delayed transfer functions to our data. More observations of other targets with this promising technique are currently underway.  

\vspace{5cm}






\bibliographystyle{aipproc}   

\bibliography{sample}

\begin{thebibliography}{00}
\bibitem{casa98}
Casares J., Charles P.A. \& Kuulkers E. 1998, ApJ, 493, L39-L42.
\bibitem{C04}
Casares J., Steeghs D., Hynes R.I., Charles P.A., Cornelisse R. \&
O'Brien K. 2004,RMxA, 20, 21.
\bibitem{C06}
Casares, J., 
Cornelisse, R., Steeghs, D., Charles, P.~A., Hynes, R.~I., O'Brien, K., \& 
Strohmayer, T.~E.\ 2006, MNRAS, 373, 1235 

\bibitem{dav75}
Davidsen, A. Margon, B., Middleditch, J. 1975, ApJ, 198, 653
\bibitem{Dhi07}
Dhillon, V.S. et al., 2007, MNRAS,378, 825 
\bibitem{gask87}
Gaskell C.M. \& Peterson B.M. 1987, ApJ, 65, 1.
\bibitem{hynes06}
Hynes, R.I., Horne, K., O'Brien,  K., Haswell, C.A., Robinson, E. L., King, A. R., 
Charles, P. A., \& Pearson, K. J., 2006, ApJ, 648, 1156
\bibitem{hynes98}
Hynes R.I., O'Brien  K., Horne K., Chen W. \& Haswell C.A. 1998,{\em MNRAS}, 299, L37.

\bibitem{midd81}
Middleditch, J., Mason, K. O., Nelson, J. E., \& White, N. E. 1981, ApJ, 244, 1001
\bibitem{MD05}
Mu\~noz-Darias T., Mart\'\i{}nez-Pais I.G. \& Casares J. 2005, in {\em INTERACTING BINARIES: Accretion, Evolution and Outcomes}, eds. L.A. Antonelli, L. Burderi, F. D'Antona, et al.
\bibitem{MD07}
Mu{\~n}oz-Darias, T., Mart{\'{\i}}nez-Pais, I.~G., Casares, J., Dhillon, 
V.~S., Marsh, T.~R., Cornelisse, R., Steeghs, D., \& Charles, P.~A.\ 2007, 
MNRAS, 379, 1637  
\bibitem{obrien02}
O'Brien K., Horne K., Hynes R.I., Chen W., Haswell C.A. \& Still M.D. 2002, MNRAS, 334, 426.
\bibitem{stee02}
Steeghs D. \& Casares J. 2002, ApJ, 568, 273-278.
\end{thebibliography}

\IfFileExists{\jobname.bbl}{}
 {\typeout{}
  \typeout{******************************************}
  \typeout{** Please run "bibtex \jobname" to optain}
  \typeout{** the bibliography and then re-run LaTeX}
  \typeout{** twice to fix the references!}
  \typeout{******************************************}
  \typeout{}
 }

\end{document}